What is affordance theory and how can it be used in communication research?

Sorin Adam Matei
Professor of Communication
Brian Lamb School of Communication
Purdue University
West Lafayette, IN 47907
smatei@purdue.edu
765 494 3666

Abstract

Affordance theory, proposed by Gibson (2014) and brought to technology research by Norman (2008), proposes that the use of an object is intrinsically determined by its physical shape. However, when translated to digital objects, affordance theory loses explanatory power, as the same physical affordances, for example, screens, can have many socially constructed meanings and can be used in many ways. Furthermore, Gibson's core idea that physical affordances have intrinsic, pre-cognitive meaning cannot be sustained for the highly symbolic nature of digital affordances, which gain meaning through social learning and use. A possible way to solve this issue is to think about on-screen affordances as symbols and affordance research as a semiotic and linguistic enterprise.

The concept of affordance starts from a simple premise: the world interacts with us just as much as we interact with it (Norman, 2008; Gibson, 2014). An affordance is what the world around us "affords" us to do with it. As beings existing in physical bodies, our lives depend on what the world gives us a chance to interact with. The inventor of the term of affordance, J.J. Gibson, proposed that the form of the objects surrounding us shape the perception of what is possible to do with them (Gibson, 2014). As we live in the physical world, we acquire perceptions of how to use the objects and features of that world. These perceptions spring out of reality as potentialities inviting us to take advantage of an object when we want to do accomplish a task; they are experiential and relational, rather than carefully thought out and discrete cognitions (Jones, 2003). The process is similar to the older concept of phenomenological "intuition," or felt meaning. According to Gibson, a flat, solid surface invites us experientially to stand or lay on by mere interaction with our feet, balance organs, and vision (Gibson, 2014). We perceive the use of that surface just as immediately as any other animal would through our sense of balance and sight; no higher cognition is involved. We do not think propositionally "there is a flat surface, let me walk on it;" we perceive experientially that the surface will support us with our feet, partially by touch and partly by the signals our inner ear send us telling us if it is possible to stand upright on it or not.

An oblong object of sufficient length to provide leverage and narrow enough to be grasped invites us to use it as a club. If threatened by an aggressor, we would grab it without any forethought. A chimpanzee would use a stick in just the same way and with just as little forethought to defend itself. The affordance of "wield-ability" and amplifying force is perceived at the most fundamental level, that of immediate response to a stimulus.

Gibson's psychology of affordances is non-conceptual, relational, and ecological (Jones, 2003). Objects compel use, and people are conditioned at the level of perception by the form, substance, or texture of the objects. In other words, objects have intrinsic, pre-cognitive meanings; they speak a language of their own, shaped by what they can do for us. Humans recognize those meanings in use, rather than adding to them meanings demanded by thought-out plans.

Created at the end of the 1930s in the narrow field of visual perception with applicability to car safety, affordance theory remained of limited interest for several decades, until the late 1960s. Only those that followed Gibson's ecological psychology as applied to visual perceptions were interested in it. Also, the theory was limited to explaining perceptions rather than to inspire broader applications. Gibson's own intention was to provide a Gestalt and phenomenological explanation of how we experience the world (Oliver, 2005). His main point was that when encountering the world, our minds do not work synthetically. They do not combine features and properties observed and qualified individually to provide a conscious plan for using them in a certain way. Our perception of what things and features of the world are or may be used for emerges in use; features are directly perceived as we interact with the world. Gibson uses the term "information pick up," a variation of the term "intuition," to describe the moment when our perception starts (Gibson, 2014).

Gibson's theory also remained an outer province of psychology because his idea that perceptions are direct and refer to the intrinsic meanings (Oliver, 2005) of objects clashed with core tenents of cognitive psychology, which claim that perceptions are cognitive processes that involve some reasoning (Tacca, 2011).

It took a former engineer converted to the psychological study of human-technology interactions, Donald Norman (Norman, 2013), to bring Gibson's relational and experiential perception psychology to the public attention, but only by mitigating the anti-cognitivist claims of Gibsonism. Working outside the realm of experimental psychology while engaged in applied research of human interactions with technology, Norman realized that Gibson's insights could be directly validated and used when designing the physical shape of products. He realized that current design practices already used affordance-thinking. His example of the push-bar door opening mechanism, which invites the user to press a door to both unlock and open it, is a canonical example. He took these examples, Gibson's psychology, and some of his insights to produce a theory of affordance-driven design.

  However, Norman rightly understood that a theory created for understanding and shaping physical objects could not be used universally as such (Norman, 2008). For example, when designing graphic user interfaces for computer applications, websites, or apps, the number of physical affordances are dramatically reduced to "looking at," "click on," "tap on," or "drag around" actions. The physical actions themselves, including something as simple as manipulating a mouse with a cursor, double-clicking, and anticipating the results, can be challenging and more important have significant cognitive loads and reasoning demands. In fact, direct perception of the meaning of very broad onscreen affordances nearly impossible. The same on-screen feature can suggest many affordances, which need to be sorted out cognitively. A click on a computer screen can produce many and different outcomes. Norman realized this very well when he proposed that on-screen interfaces propose "perceived affordances" (Norman, 2013), which are subjective, involve some learning, and can be quite numerous. However, this is a poor choice of terms, which did not consider the rigorous definition of perception proposed by Gibson, according to whom, affordance perception is immediate and direct. If according to Gibson, affordances are direct perceptions (Gibson, 1977), the "perceived" modifier added by Norman to "affordances" is both redundant and confusing. Replacing the terms in the equation "online

affordances = perceived affordances," with those suggested by the equation "affordances = perceptions," we obtain "online affordances = perceived perceptions," which is rather nonsensical. However, Norman uses the terms "perceived" in a far less prescriptive manner than Gibson. For him, perceived is akin to "supposed" or "inferred." "Perceived" is a mere synonym for any modifier that would turn "affordances" from "something-that-is-demanded-by-use" into "something-that-I-infer-this-thing-can-do." Norman shifts the term from essential characteristics to assumed potential. An affordance does not have intrinsic meaning; the meaning is constructed cognitively by the user. For Gibson, affordances are "invariant characteristics," for Norman "reasoned possibilities for action." Furthermore, Norman emphasizes that affordances should be visible and understandable, while for Gibson, affordances may exist in situations where visibility is not necessary.

Oliver (Oliver, 2005) observes that recast in this manner, affordances tend to become signals. In fact, Norman himself suggests that affordances are elements of communication, whose role is to indicate where and what the user can do in a given situation. This resignification of the affordance concept turns, in Oliver's words, the designer into an author and the user into a reader. The conclusion of this argument is that researching human on-screen or online experiences should be considered as a subfield of semiotics, not psychology. Affordances become symbols that steer action through meanings accumulated in time and learned. Affordances are now less experiential and more conventional. They refer to mental maps for organizing the world of interactions and handling objects. Affordances are heavily cognitive, are liable to many interpretations, and are influenced by context.

The conclusion is that on-screen "affordances" are metaphoric and they belong, research-wise, to communication inquiry. Understanding what people see, what they think it is possible to do, and what they do is ruled by a lexical system with nouns (features) that have socially understood meanings (learned potentialities) and verbs (actions or behaviors) that emerge from past experience and through cultural learning. We can thus say that on-screen affordances are in fact symbols. They have semantic meaning, are concatenated by syntactic rules and by a simple grammar that defines what symbol impels the user to do in a certain situation. While the idea of affordances remains a powerful tool that may inspire designers to think in a grounded way about user-centric interface design, so that the users engage technologies organically, academic researchers need to think about them semiotically. Each online or on-screen feature affords an action as a consequence of a meaning attached to it by the users based on her past experience. Meaning results from metaphoric transfer of significance and potential use from cognate domains (Lakoff & Johnson, 2003), or by observing and communicating with other users. In a way, affordance research ends up being a native communication problem, and in this respect, we, communication scholars, have much to benefit from recasting the problem this way.